\documentclass[conference]{IEEEtran}
\def\ps@headings{%
\def\@oddhead{\mbox{}\scriptsize\rightmark \hfil \thepage}%
\def\@evenhead{\scriptsize\thepage \hfil \leftmark\mbox{}}%
\def\@oddfoot{}%
\def\@evenfoot{}}
\makeatother
\usepackage{booktabs} 
\usepackage{url}
\usepackage{svg}
\usepackage{etoolbox}
\usepackage{cuted}
\usepackage{lipsum}
\usepackage{amssymb}
\usepackage{array}
\usepackage{amsmath}
\usepackage{tikz}
\usepackage{color}
\usepackage{dirtytalk}
\usepackage{algpseudocode}
\usepackage{algorithm}
\usepackage{graphicx}
\usepackage{caption}
\usepackage{mdframed}
\usepackage{tcolorbox}

\ifCLASSINFOpdf
\else
\fi
\DeclareRobustCommand*{\IEEEauthorrefmark}[1]{\raisebox{0pt}[0pt][0pt]{\textsuperscript{\footnotesize #1}}}

\begin{document}
%
\title{Access Control for Electronic Health Records with Hybrid Blockchain-Edge Architecture}


\author{\IEEEauthorblockN{Hao Guo\IEEEauthorrefmark{1}~~~~Wanxin Li\IEEEauthorrefmark{2}~~~~Mark Nejad\IEEEauthorrefmark{2}~~~~Chien-Chung Shen\IEEEauthorrefmark{1}}
\IEEEauthorblockA{\IEEEauthorrefmark{1}Department of Computer and Information Sciences\\\IEEEauthorrefmark{2}Department of Civil and Environmental Engineering\\
University of Delaware, U.S.A. \\
\
\{haoguo,wanxinli,nejad,cshen\}@udel.edu}}


%


\maketitle

\begin{abstract}
The global Electronic Health Record (EHR) market is growing dramatically and expected to reach \$39.7 billions by 2022. To safe-guard security and privacy of EHR, access control is an essential mechanism for managing EHR data. This paper proposes a hybrid architecture to facilitate access control of EHR data by using both blockchain and edge node. Within the architecture, a blockchain-based controller manages identity and access control policies and serves as a tamper-proof log of access events. In addition, off-chain edge nodes store the EHR data and apply policies specified in Abbreviated Language For Authorization (ALFA) to enforce attribute-based access control on EHR data in collaboration with the blockchain-based access control logs. We evaluate the proposed hybrid architecture by utilizing Hyperledger Composer Fabric blockchain 
to measure the performance of executing smart contracts and ACL policies in terms of transaction processing time and response time against unauthorized data retrieval.
\end{abstract}

\begin{IEEEkeywords}
Attribute-based Access Control, Access Control List, Blockchain, Edge Computing, Hyperledger, Smart Contract.
\end{IEEEkeywords}


%
\IEEEpeerreviewmaketitle

\section{Introduction}
In eHealth, data for the Electronic Health Records (EHRs) of patients can be gathered from multiple sources, such as wearable devices, smart sensors, and medical imaging equipment. It has been reported that the amount of EHR data will continue to grow at a rate of {48\%} each year to reach 2,314 zettabytes by 2020~\cite{ehrgrow}. However, according to the U.S. Department of Health and Human Services, there were more than 2,181 cases of healthcare data violations between 2009 and 2017, resulting in the exposure of 176,709,305 medical records~\cite{medrecords}. As a result, safe-guarding the EHR data has become a pivotal issue in eHealth. Although encryption addresses some fundamental security and privacy issues of EHR, access control, in particular, is difficult to enforce effectively due to the highly distributed and fragmented nature of EHR data and the complex relationship between data owners and data users. Therefore, providing a flexible and fine-grained access control solution for EHR data is of paramount interest.

Recently, blockchain has been suggested to be a promising solution for EHR data management~\cite{kamau2018blockchain}. The inherent secure-by-design feature of a blockchain-based infrastructure has the potential to provide a tamper-proof log for all the access events of EHR. In particular, all the access events can be verified and recorded through a consensus mechanism before being added to the blockchain. However, from the prospective of EHR management, the traditional blockchain-based solutions suffer from two significant drawbacks. First, although blockchain can ensure data integrity, it lacks proper access control mechanisms to contain operations performed by different participants.
Second, the size of blocks in a blockchain is too limited to accommodate EHR data containing images (e.g., X-ray, CT scan, and MRI) and/or videos (e.g., ultrasound).



This paper proposes a hybrid architecture of using both blockchain and edge nodes to facilitate attribute-based access control of EHR data.
Specifically, the Hyperledger Composer Fabric~\cite{hyperledgerfabric} blockchain executes smart contracts programmed with Access Control Lists (ACLs) to enforce identity-based access control of EHR data and log legitimate access events into blockchain for traceability and accountability. 
In collaboration, edge nodes store EHR data and further enforce attribute-based access control (ABAC)\footnote{ABAC is formally defined as {\em an access control method where subject requests to perform operations on objects are granted or denied based on assigned attributes of the subject, assigned attributes of the object, environment conditions, and a set of policies that are specified in terms of those attributes and conditions} \cite{nist}.} of EHR data with policies specified in the Abbreviated Language For Authorization (ALFA)~\cite{wiki:alfa}. ALFA maps directly into eXtensible Access Control Markup Language (XACML) and provides succinct representation.
In addition, hash digest is used to protect the integrity of EHR data stored in the edge nodes, which helps detect any alteration of EHR. Furthermore, one-time self-destructing urls~\cite{1ty}, containing the addresses of EHR data on the edge nodes, are referenced in smart contracts, which will be returned to the healthcare providers after the successful execution of the ACL access policy. The healthcare providers then use the urls to access EHR data from edge nodes. Therefore, only eligible users who pass the attribute-based access control imposed by edge nodes can access the requested EHR data.

To validate our design, we prototype the hybrid architecture by using the Hyperledger Composer Fabric platform. In addition, we conduct multiple experiments to validate both smart contracts and access control policies which show that the proposed system can maintain a traceable access events and transaction records for EHR data management. We evaluate the system performance, via multiple experiments, of the transaction processing time and the average response time against unauthorized EHR data request under different settings.

The remainder of this paper is organized as follows. Section \ref{pre} reviews the backgrounds of attribute-based access control with XACML and ALFA, edge node with eHealth sensor platform, and Hyperledger Composer Fabric platform. The proposed hybrid architecture of using both blockchain and edge nodes to facilitate ABAC policies of EHR data and the detailed designs of ACL policies are presented in Section \ref{arch}. In Section \ref{evaluation}, we conduct experiments to test access control mechanisms and the performance of the system. In Section \ref{security}, we analyse the proposed scheme with respect to potential attacks. Related work is described in Section \ref{related}, and Section \ref{conclusion} concludes the paper with future research directions.

\section{Preliminary}
\label{pre}

In this section, we review the eXtensible Access Control Markup Language (XACML), the Abbreviated Language For Authorization (ALFA), eHealth sensor platform and edge node, and the Hyperledger Composer Fabric platform.


\subsection{XACML and ALFA}

XACML is a standard that defines a fine-grained, attribute-based access control policy language. XACML also defines a reference architecture and a processing model to evaluate access requests according to the rules defined in access policies. 

\begin{figure}[h]
\centering
\includegraphics[width=0.483\textwidth]{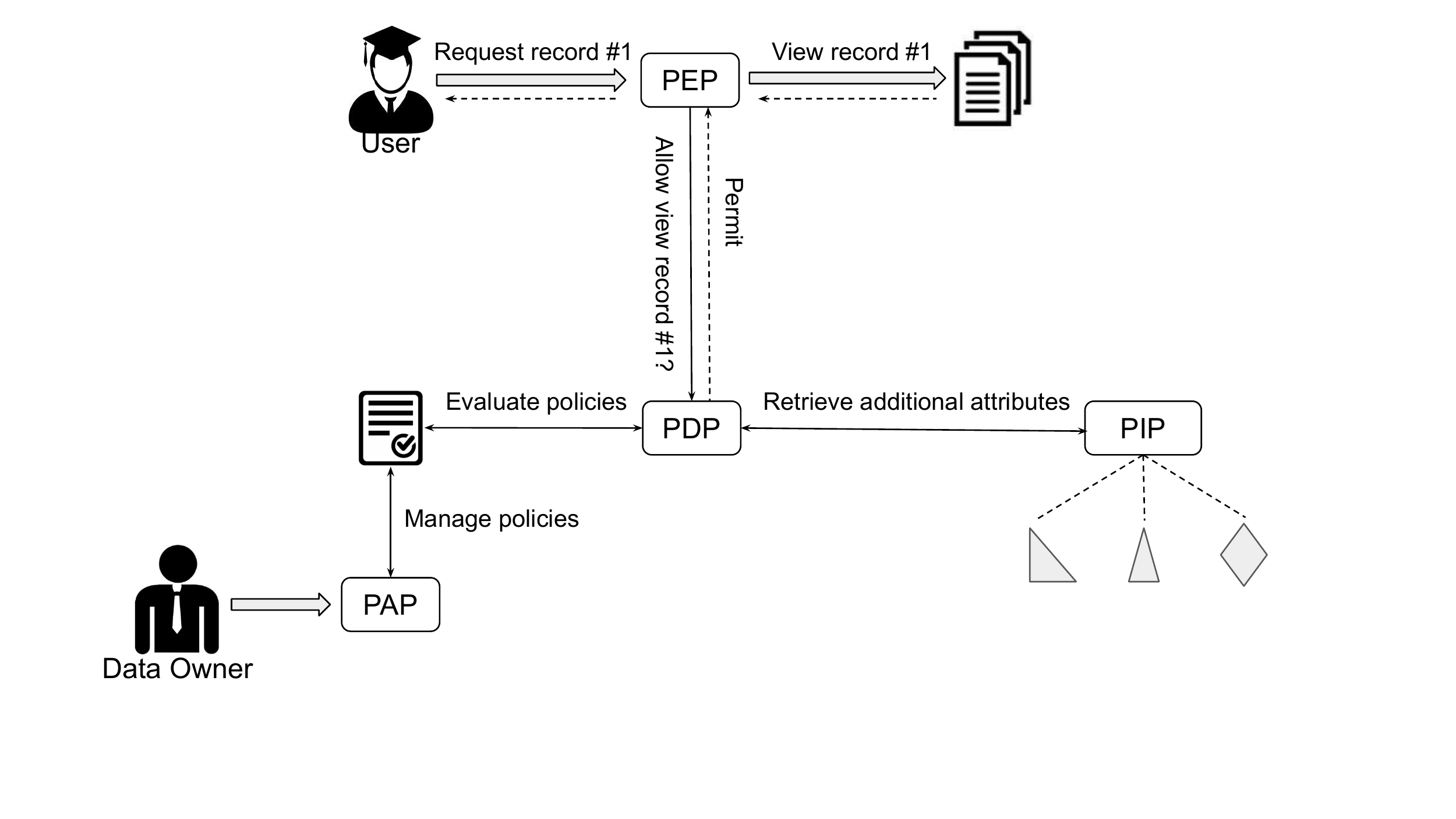}
\caption{XACML architecture and processing model.}
\label{fig:xacml}
\end{figure}

As shown in Fig. \ref{fig:xacml}, the architecture of XACML consists of the following components.
The Policy Enforcement Point (PEP) inspects a user's access request, converts the request into an XACML authorization request and forward it to the Policy Decision Point (PDP) to obtain an access decision (e.g., permit or deny), and acts on the received decision. The Policy Administration Point (PAP) is responsible for managing access control policies.
PDP evaluates incoming requests against policies it has been configured with and returns access decision (approve or deny). PDP may also use Policy Information Point (PIP) to retrieve attribute values.
PIP connects PDP to retrieve external sources such as source of attribute values (i.e. a resource, subject, environment).


However, due to the verbose design pattern and complicated syntax of XACML, many developers find it difficult to write and read XACML scripts. As a result, we use ALFA, which can specify access control policies and be mapped directly into XACML structures.
ALFA contains the same elements as XACML defined and inherit the overall structure and concepts of XACML. 
The key components of ALFA are: Attributes, Target, Conditions, Rule, Policy, and Obligation. We use ALFA to enforce ABAC policies for EHR data stored in edge nodes. By utilizing above components of ALFA, we can impose various ABAC policies on EHR data, based on data owner's specification to provide a comprehensive decision process.

\subsection{eHealth Sensor Platform and Edge Node}

\begin{figure}[h]
\centering
\includegraphics[width=0.41\textwidth]{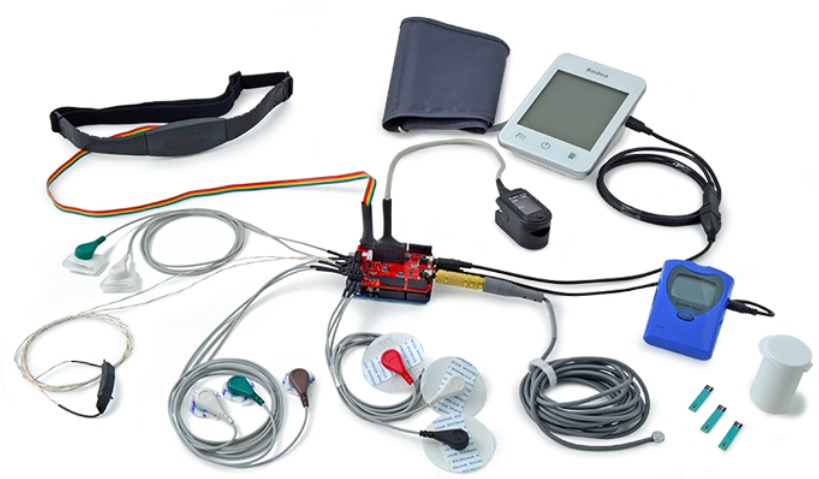}
\caption{eHealth sensor platform.}
\label{fig:sensorplatform}
\end{figure}

\begin{figure}[h]
\centering
\includegraphics[width=0.45\textwidth]{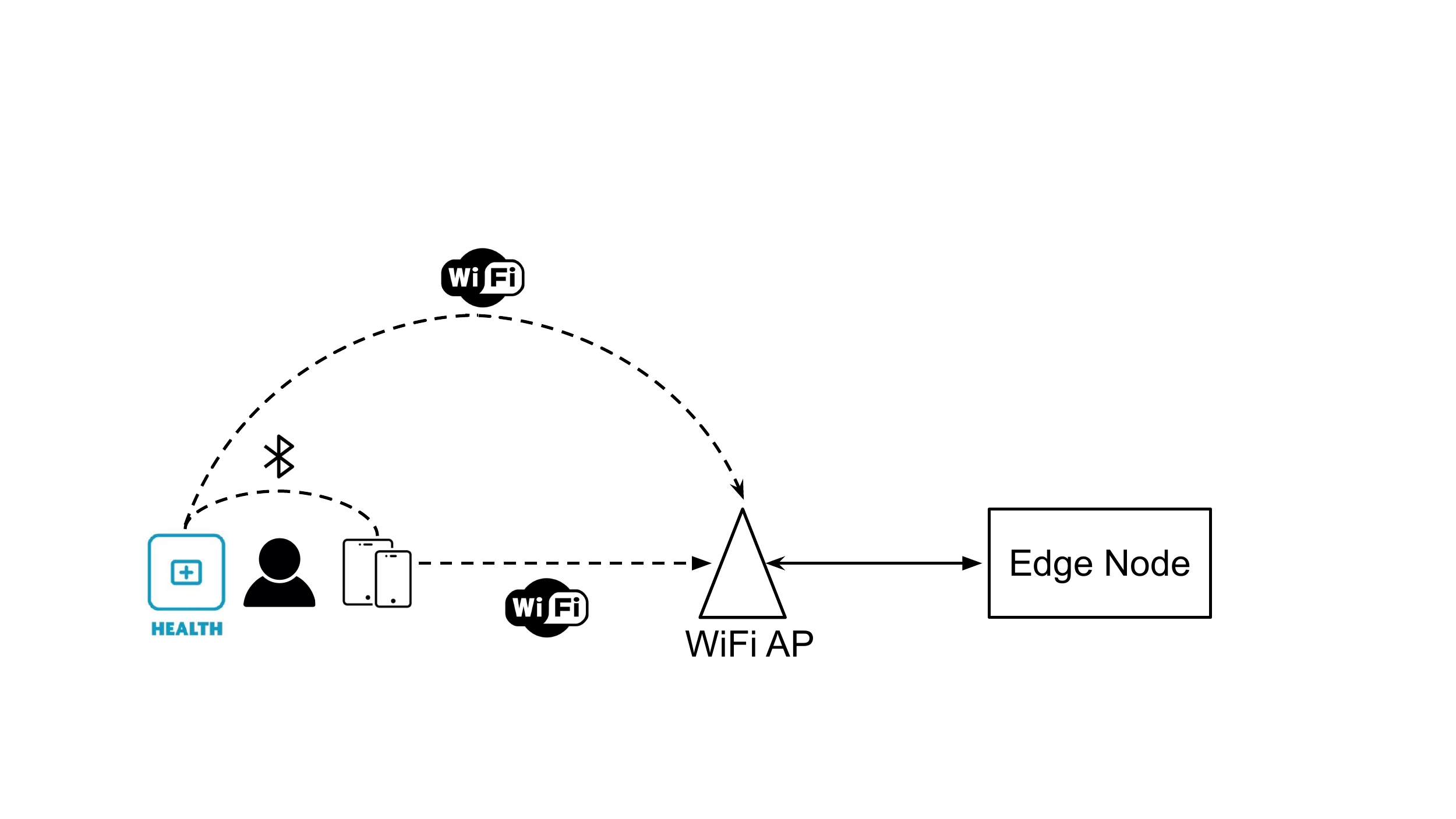}
\caption{Data collection from sensors to an edge node.}
\label{fig:sensordata}
\end{figure}

Recent advancements in smart wearable sensor technologies enable EHR systems to collect patient's bio-metric data. For example, the MySignals~\cite{mysignal} eHealth sensor platform, shown in Fig. \ref{fig:sensorplatform}, measures 16 different bio-metric parameters via different sensors, such as blood pressure sensor, airflow sensor, Electrocardiogram sensor (ECG), Electromyography sensor (EMG), snore sensor, body position sensor, and Galvanic Skin response sensor (GSR). Such platform may also upload EHR data directly to an edge node via WiFi or communicate the EHR data to patient's smartphone via Bluetooth, which uploads the EHR to the edge node, as depicted in Fig. \ref{fig:sensordata}. 

\subsection{Hyperledger Composer Fabric Platform}
Hyperledger~\cite{hyperledger} is an open source blockchain project started in December 2015 by the Linux Foundation and supported by industrial companies such as IBM, Intel and SAP. Hyperledger Composer Fabric~\cite{hyperledgerfabric} is a blockchain implementation platform providing a modular and extensible architecture that allows components, such as consensus mechanism, to be plug-and-play. In addition, it leverages the container technology to host smart contracts, termed ``chaincode,'' that comprise the application logic of the system~\cite{hyperledgerfabric}. 

Hyperledger Composer Fabric supports chaincode written in Go and JavaScript out-of-the-box, and other languages such as Java by installing appropriate modules~\cite{hyperledger}. As a result, Hyperledger Composer Fabric is more flexible than the Ethereum blockchain platform which only support a closed smart contract language named Solidity. Compared with Ethereum's public and permissionless blockchain network, Hyperledger provides a private and permissioned network which enables that only  privileged entities and nodes can participate in the network. Moreover, 
Hyperledger has a modular architecture and provides a lot of flexibility in how you want to use it. 

\begin{figure*}[ht]
\centering
\includegraphics[width=0.79\textwidth]{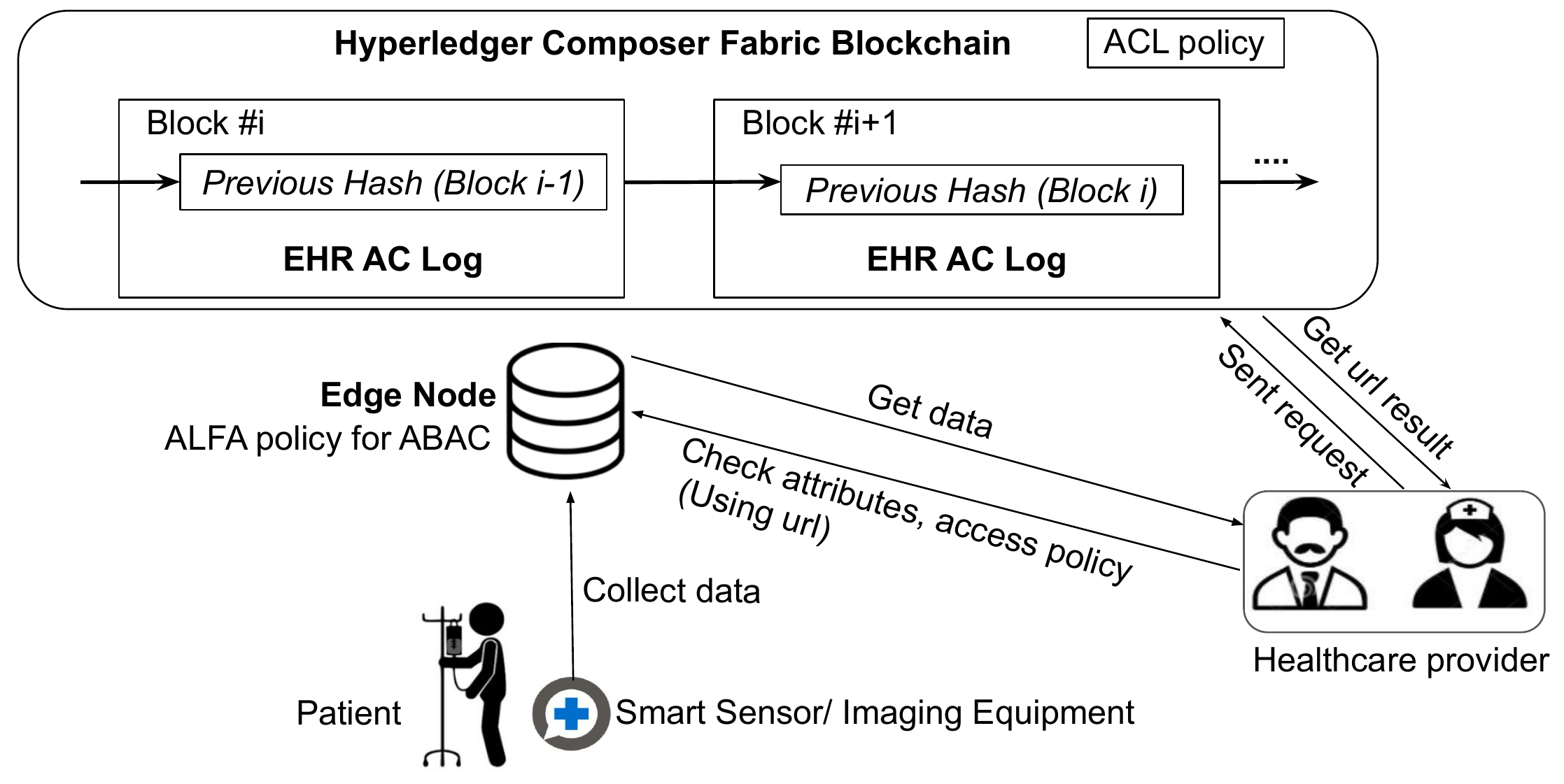}
\centering
\caption{Hybrid architecture of blockchain and edge node.}
\label{fig:framework}
\end{figure*}

\section{System Architecture}
\label{arch}

In this section, we illustrate the hybrid architecture to facilitate access control of EHR data by using both blockchain and edge node. By referring to Fig. \ref{fig:framework}, we first enumerate the following entities which take part in the architecture.

\begin{itemize}
\item Patient: A patient is an entity who owns the EHR data to be accessed. A patient may specify the access policies for the EHR data he/she owns.

\item Healthcare provider: A healthcare provider (e.g., doctor and nurse) is an entity who needs to access EHR data owned by patients. A healthcare provider actively seeks access authorizations from patients.

\item Smart sensor/imaging equipment: A smart sensor is a device which collects EHR data from patients and sends it to the edge node. Imaging equipment may include X-ray, CT scan, MRI, and ultrasound, which generate EHR data from patients.

\item EHR data: An EHR data is a piece of information owned by a patient, and can be accessed by authorized healthcare providers.

\item Edge node: An edge node is a computing and storage device which stores EHR data and imposes attribute-based access control policies.  

\item Blockchain: Blockchain is utilized as the controller of the architecture which manages access control policies and serves as a tamper-proof access log.
\end{itemize}

\begin{figure*}[ht]
\centering
\includegraphics[width=0.90\textwidth]{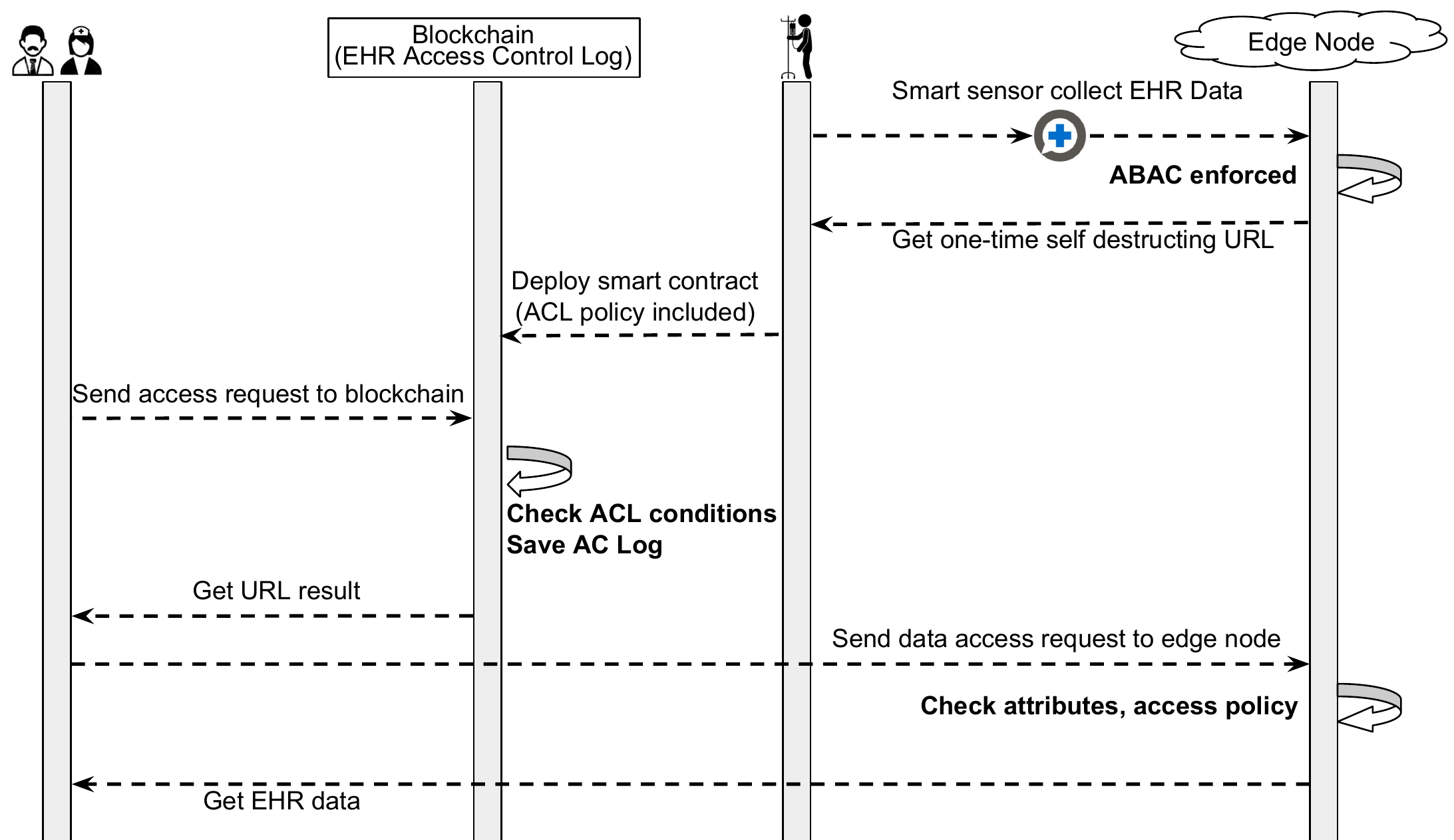}
\centering
\caption{Access control workflow.}
\label{fig:workingflow}
\end{figure*}

Access control workflow on the architecture is depicted as Fig. \ref{fig:workingflow}. First, smart sensors and imaging equipment collect EHR data from the patient and upload it to the edge node. Second, the edge node applies ABAC policies to enforce the access permissions for the EHR data and returns a one-time self-destructing url to the patient which contains the address of the EHR on the edge node. Third, the same patient will register with the Hyperledger Composer Fabric blockchain  and define the Access Control List (ACL) policy to declare access permissions to the healthcare providers. Next, Doctors/nurses can send the access request through smart contract, and it will check the identity information against ACL policy. If the condition has been satisfied, the smart contract will return the corresponding url address to locate the edge node which stores the EHR data separately. Finally, Doctor or nurses can get access to EHR data as long as they satisfy the requirement of ABAC policies enforced on the EHR data.

In the remainder of this section, we first illustrate the Hyperledger Composer Fabric blockchain network and describe the design of ACL policy. Next, we describe the hash digest mechanism for EHR data and one-time self-destructing urls.

\subsection{Hyperledger Composer Fabric Blockchain Network and Access Control List with Smart Contract}

\begin{figure}[h]
\centering
\includegraphics[width=0.457\textwidth]{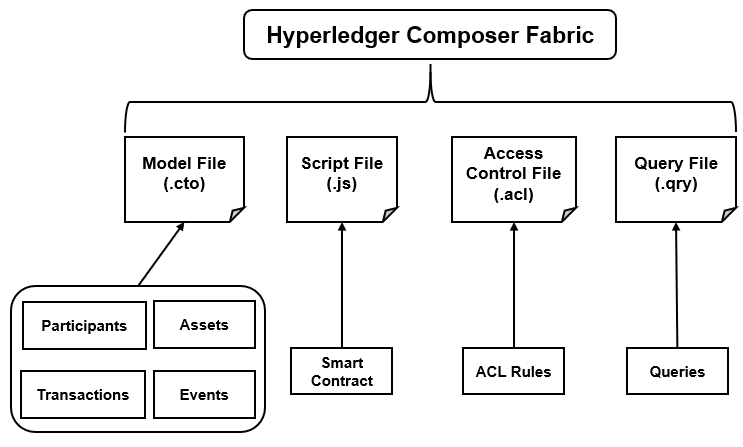}
\caption{Hyperledger Composer Fabric infrastructure.}
\label{fig:hyperledger}
\end{figure}

To implement and evaluate the access control mechanism, we designed and developed a blockchain-based web application on the Hyperledger Composer Fabric. As shown in Fig. \ref{fig:hyperledger}, the Hyperledger Composer Fabric infrastructure contains four programmable parts. The first part is to define the network features in a Model File (.cto), which includes participants, url addresses, transactions, and an access control log. The second part is to write smart contracts in the Script File (.js), which contain transaction processing functions. The third part is to define ACL rules for different participants in the Access Control File (.acl). Finally, database queries are defined in the Query File (.qry).

Hyperledger Composer Fabric provides a developing platform for blockchain-based applications, such as cryptocurrencies. However, a cogitative component in the design is still needed to transform the Blockchain network from cryptocurrencies to our proposed system. In a classic cryptocurrency architecture, participants represent people or organizations who take part in the business network. Authorized by ACL, participants are allowed to transact assets (e.g., Bitcoins) among the network. Above all, business logic for transactions are defined in smart Contract. In this architecture, blockchain network is still limited to digital-currency use case. 

In our proposed system, participants represent patients and doctors. Patient’s EHR address is saved as personal asset in the blockchain network. In other words, patient and EHR asset have one-to-one pair relationship, which are both identified by patient ID. Authorized by ACL, a doctor is allowed to retrieve his patient’s EHR addresses while the patient is only allowed to retrieve his or her own EHR address. The retrieving function is defined as transaction process in the smart contract and invoked by participants submitting their requests. Finally, all historical retrieving events are saved as immutable and traceable EHR access control log on the blockchain network.

After the edge nodes collect the EHR data from the patient, the patient can enforce the ACL policy on his or her EHR data. By defining the ACL policies, it can determine which data users are allowed to read, write, and update the content. When the ACL policy is available in blockchain network, data users such as a doctor or a nurse can send the access request to the patient to get access approval and receive the url address for the actual EHR data stored in the edge node. 

An ACL policy is defined with the follows components:
\begin{itemize}
\item Subjects: It defines the person or entity which is involved in the ACL procedure. 

\item Operations: It indicates the action which the rule governs. In our scheme, three actions are supported: READ, WRITE, and UPDATE. 

\item Objects: It defines the objects which the ACL rules applies to. It can be either a single document of EHR data or a complicated union of EHR data. 

\item Conditions: It is an AND-gate policy expression over multiple variables. In addition, our scheme can support if(...) expression for the complex condition of ACL rules.

\item Actions: It indicates the final action of the ACL rules. It must be either ALLOW or DENY.
\end{itemize}

We define two types of ACL rules: non-conditional and conditional. Non-conditional rules are used to control access policy to a specific group of participants. On contrast, conditional ACL rules can express various AND-gate access control policy and return the Boolean result on the action result. For instance, the rule below states that only the doctors from the Christiana Hospital can READ the EHR data from patients:
\begin{verbatim}
rule Rule1 {
  description: "Only doctor from the
   Christiana Hospital can read data"
  subject(v): "Christiana.Doctor"
  operation: READ
  object(t):"Christiana.patient#123.data"
  condition: "v.role == Doctor &&
   v.organization == Christiana"
  action: ALLOW
}
\end{verbatim}
The use of ACL policy provides the access control log of events to EHR data, determining the ability to read, write or update. In another words, we provide the conditions and restrictions as to: `who' has the ability to do `what' on the blockchain ledger.

\subsection{Hash Digest for EHR Data}

 When the edge node sends the url results back to the patient, it includes the hash digest result for the EHR data. The hash digest result contains a string of digits created by a one-way hashing formula. The hash digest function can protect the integrity of EHR data and detect changes or alterations to any part of the data. Comparing the hash digests result, people can determine whether any changes have been made on EHR data. If EHR data has been modified, the hash digest is different from the original one.

\begin{figure}[h]
\centering
\includegraphics[width=0.46\textwidth]{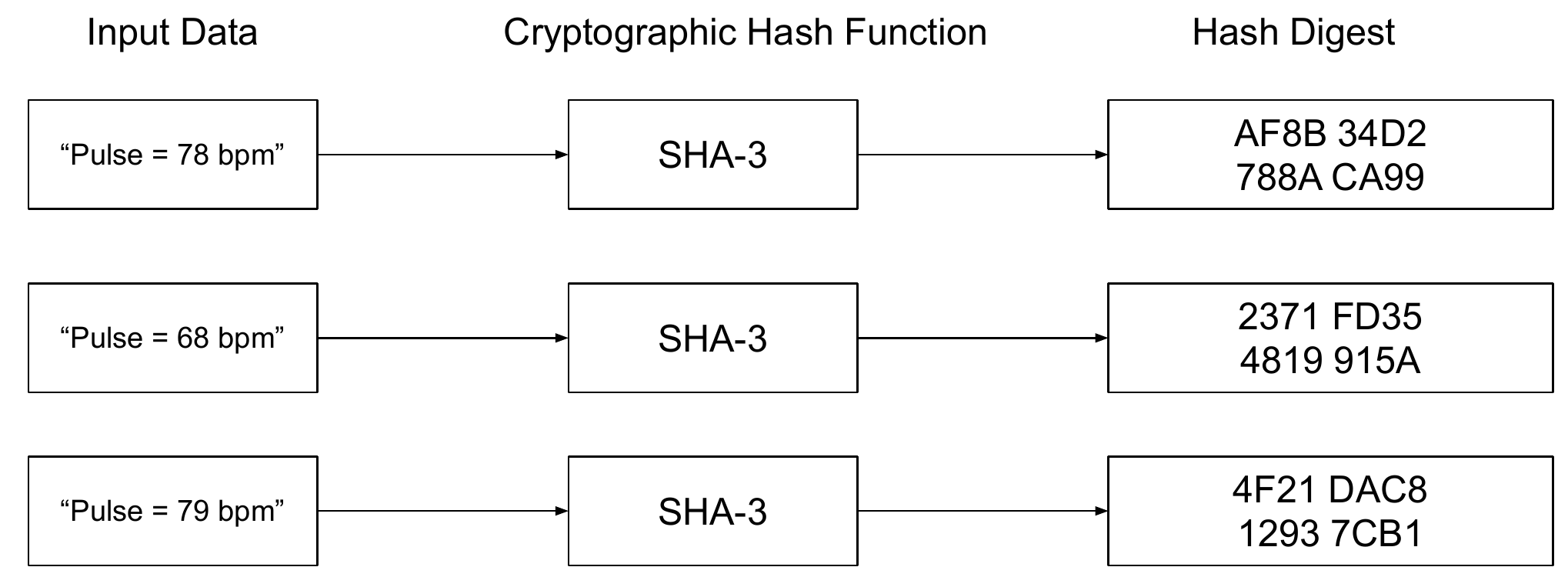}
\caption{A SHA-3 hash function at work.}
\label{fig:sha3work}
\end{figure}

Secure Hash Algorithm 3 (SHA-3) is the latest member of Secure Hash Algorithm standard released by NIST on August 2015~\cite{sha3}.
By including the hash digest result into the smart contract and transaction, that piece of information is forever saved on the blockchain network. For instance, in Fig. \ref{fig:sha3work} the original EHR saves the information as ``Pulse = 78 bpm'' and it generates the corresponding hash digest. If later someone maliciously alters the EHR as ``Pulse = 68 bpm'' or ``Pulse = 79 bpm'', the system will automatically generate a new hash digest indicating that the EHR data has been tampered.

\subsection{One-time Self-Destructing URLs}

After a doctor sends an access request to the blockchain network, the enforced ACL policy will check the information provided by the doctor. If the identity information satisfies ACL policy, the smart contract executed between the patient and the doctor saves the access log into the blockchain network and returns a one time self-destructing url to the address of EHR data.

We use {\tt https://1ty.me/}~\cite{1ty} to contain the address of EHR data stored in edge nodes. {\tt https://1ty.me/} runs on HTTPS and the EHR data address information is encrypted. The key to decrypt the address is a portion of data which is contained in the url. This generated url is not stored on the {\tt https://1ty.me/} server. As a result, only the valid and one-time url link can display and decrypt the address information. Once the address information is viewed, the encrypted information will be removed from the system and the url link will vanish and cannot be accessed again. 

\section{Experiments and Evaluation}
\label{evaluation}

\subsection{Experimental Setup}
\begin{figure}[h]
\centering
\includegraphics[width=0.48\textwidth]{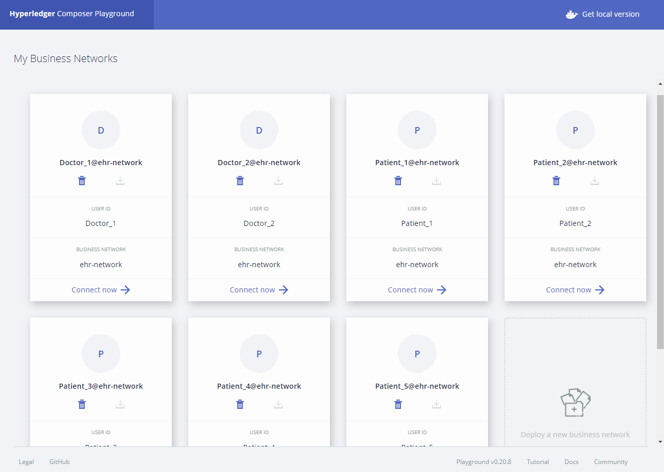}
\caption{Blockchain-based EHR network login window.}
\label{fig:login}
\end{figure}

We conduct experiments to evaluate the performance of the proposed hybrid access control system prototyped on the Hyperledger Composer Fabric framework.  Fig. \ref{fig:login} shows an example experiment with two doctors and five patients as participants, where each participant has a unique digital ID card to log into the blockchain network.  The first three patients scheduled appointments with doctor \#1 and the remaining two patients scheduled appointments with doctor \#2.  We tested the access control mechanism by logging into the prototyped blockchain network on a Chrome browser, submitting transactions, and recording access events and results of the participants. Specifically, we compared the system response time of the systems with different numbers of patients to measure the performance of the proposed system. The experiments were conducted on a computer with a 2.9 GHz Intel i5 processor, 8GB of memory, and 60 Mbps of Ethernet connection as the default configuration.  

\subsection{Experimental Results}

\subsubsection{Test of Access Control on Doctor}
\begin{figure}[h]
\centering
\includegraphics[width=0.485\textwidth]{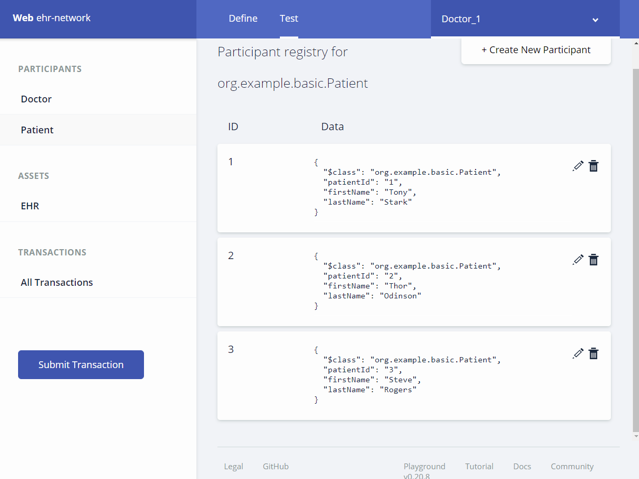}
\caption{A doctor accesses his/her list of patients.}
\label{fig:doctor}
\end{figure}

\begin{figure}[h]
\centering
\includegraphics[width=0.48\textwidth]{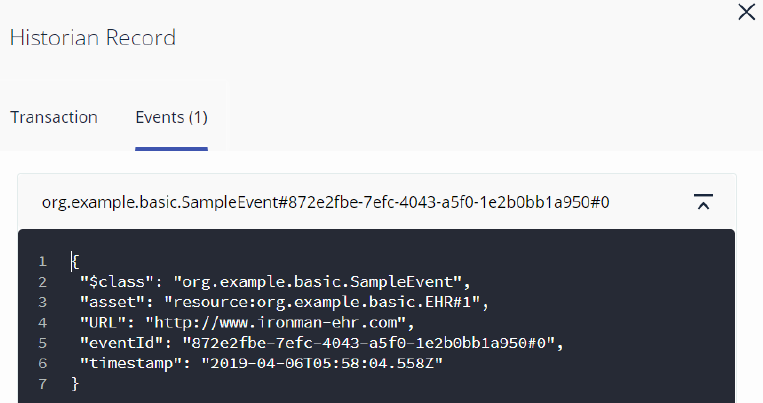}
\caption{A doctor can retrieve his/her patients' EHRs.}
\label{fig:doctorview}
\end{figure}

In the first experiment, we log into the blockchain network as a doctor. A doctor is allowed to access his/her list of patients, including ID numbers, first and last names, as shown in Fig. \ref{fig:doctor}. If a doctor wants to retrieve one specific patient's EHR data, the doctor needs to submit a transaction request by entering the patient's ID number. Upon submitting the request, the system will return the URL address of the patient's EHR. At the same time, the blockchain network will permanently record this retrieval action as a transaction event, including the event ID and the timestamp as shown in Fig. \ref{fig:doctorview}. Any EHR retrieval record can be traced back in real-time for analysis.

\subsubsection{Test of Access Control on Patient}

\begin{figure}[h]
\centering
\includegraphics[width=0.48\textwidth]{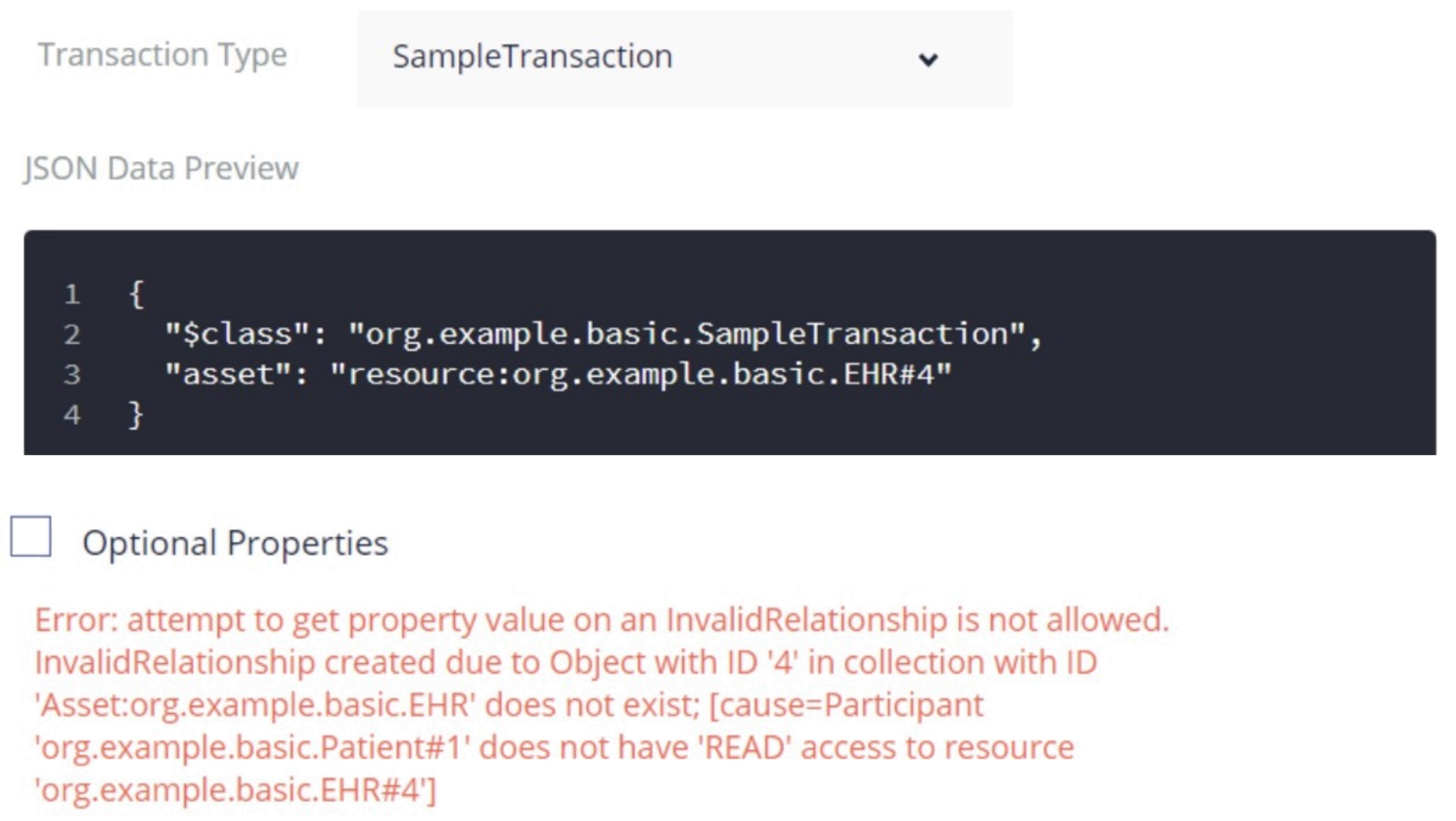}
\caption{A patient cannot retrieve other patients' EHRs.}
\label{fig:error}
\end{figure}

In the second experiment, we change the login user to a patient who can retrieve his or her own EHR data. Different from the doctors, a patient is not allowed to read other patients' information including the names and EHRs. For instance, if patient Tony Stark ({\em patient ID}: 1) attempts to retrieve the EHR data of patient David Banner ({\em patient ID}: 4), the system will reject this transaction request and return an error message indicating that this request is not allowed, as shown in Fig. \ref{fig:error}.

\subsubsection{System Performance}

For each scenario, we conducted four rounds of tests to measure the average transaction processing time and the average response time against unauthorized retrieval, which were 40 ms and 30 ms, respectively. The fact that our prototype can produce results in tens of milliseconds makes the architecture suitable to be incorporated in practical EHR access control systems. In a recent study~\cite{li2019blockchain}, Li et al. showed that the performance of Hyperledger Composer Fabric can be affected by the processor clock speed because each participant can run the same code and save the same data in a distributed architecture on different computers with different specifications and workloads. Therefore, the transaction processing time can slightly vary based on users' devices.

\begin{figure}[h]
\centering
\includegraphics[width=0.455\textwidth]{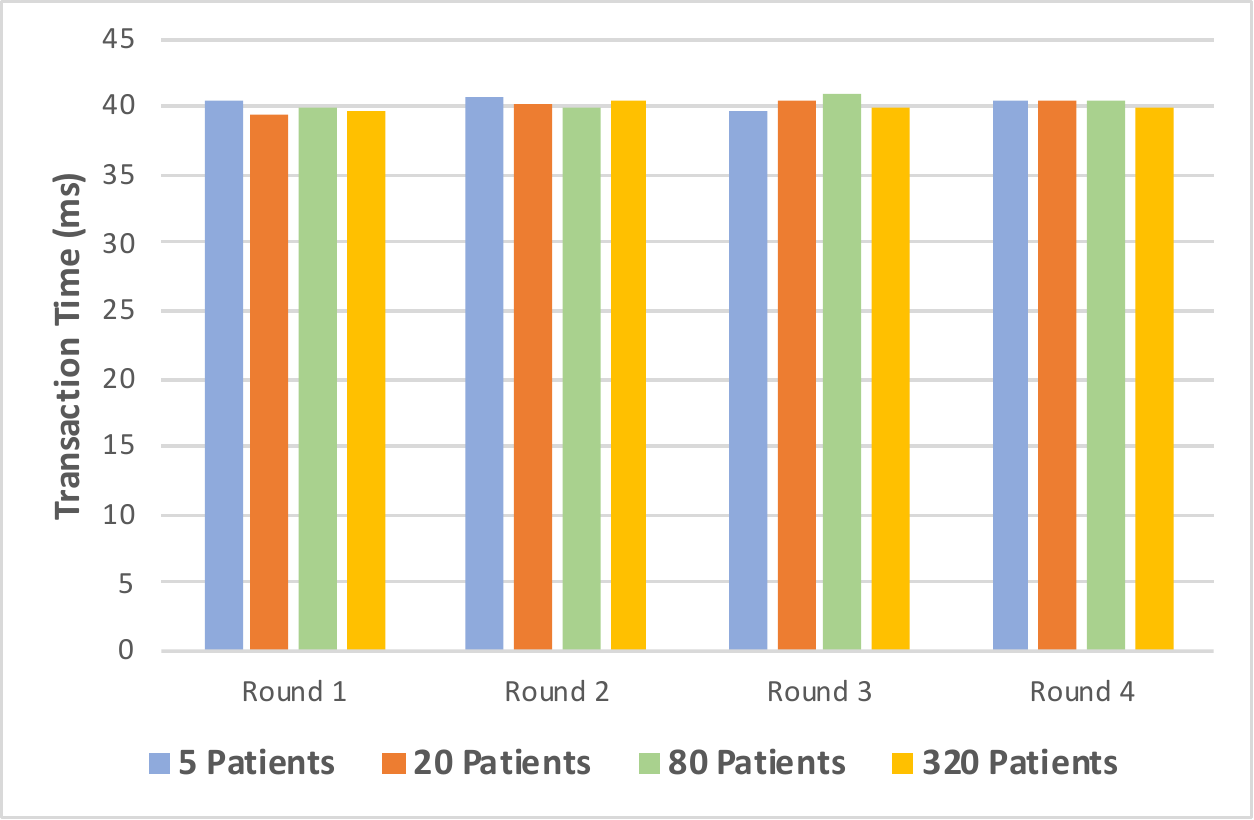}
\caption{Transaction processing time.}
\label{fig:transactiontime}
\end{figure}

\begin{figure}[h]
\centering
\includegraphics[width=0.455\textwidth]{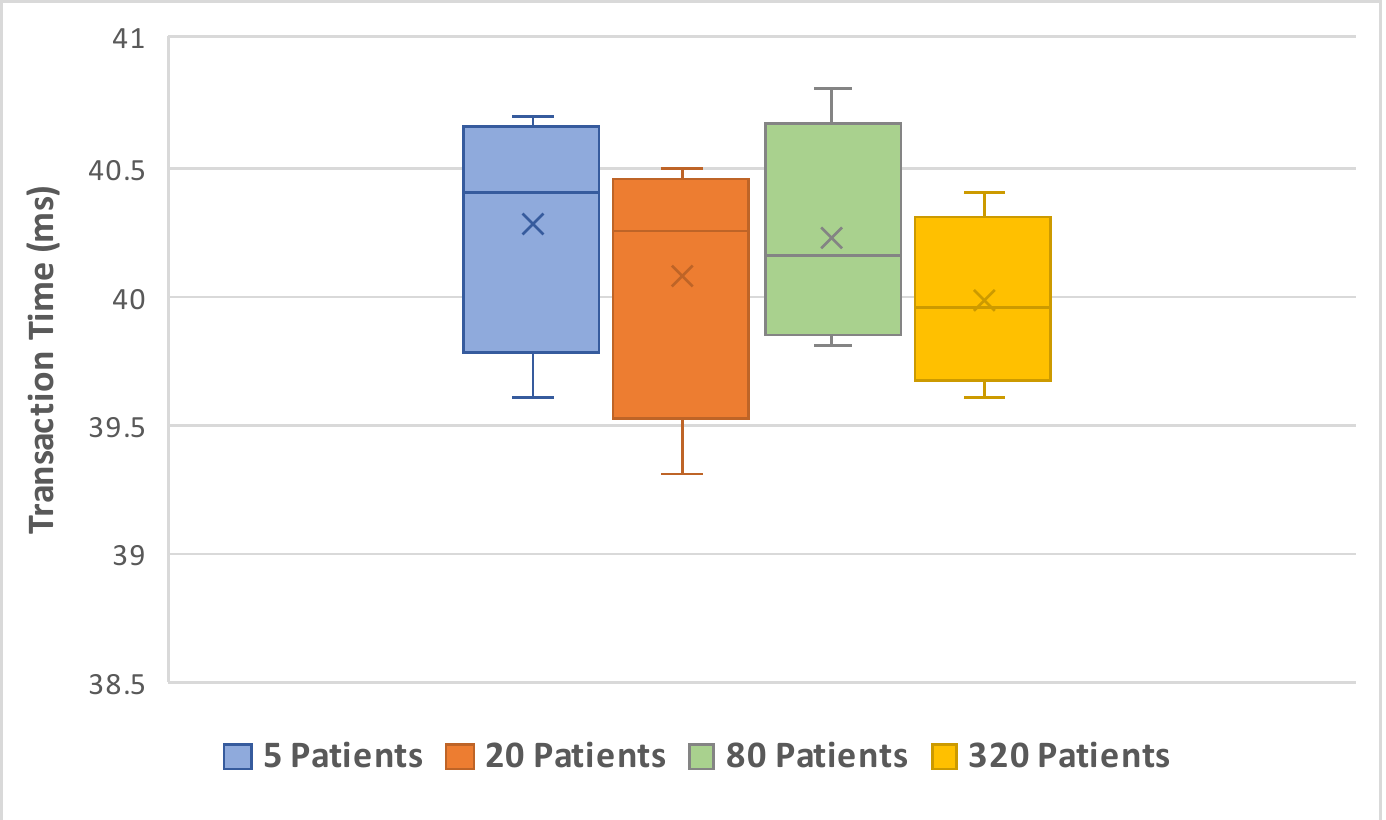}
\caption{Average transaction processing time.}
\label{fig:transactiontime_box}
\end{figure}

\begin{figure}[h]
\centering
\includegraphics[width=0.455\textwidth]{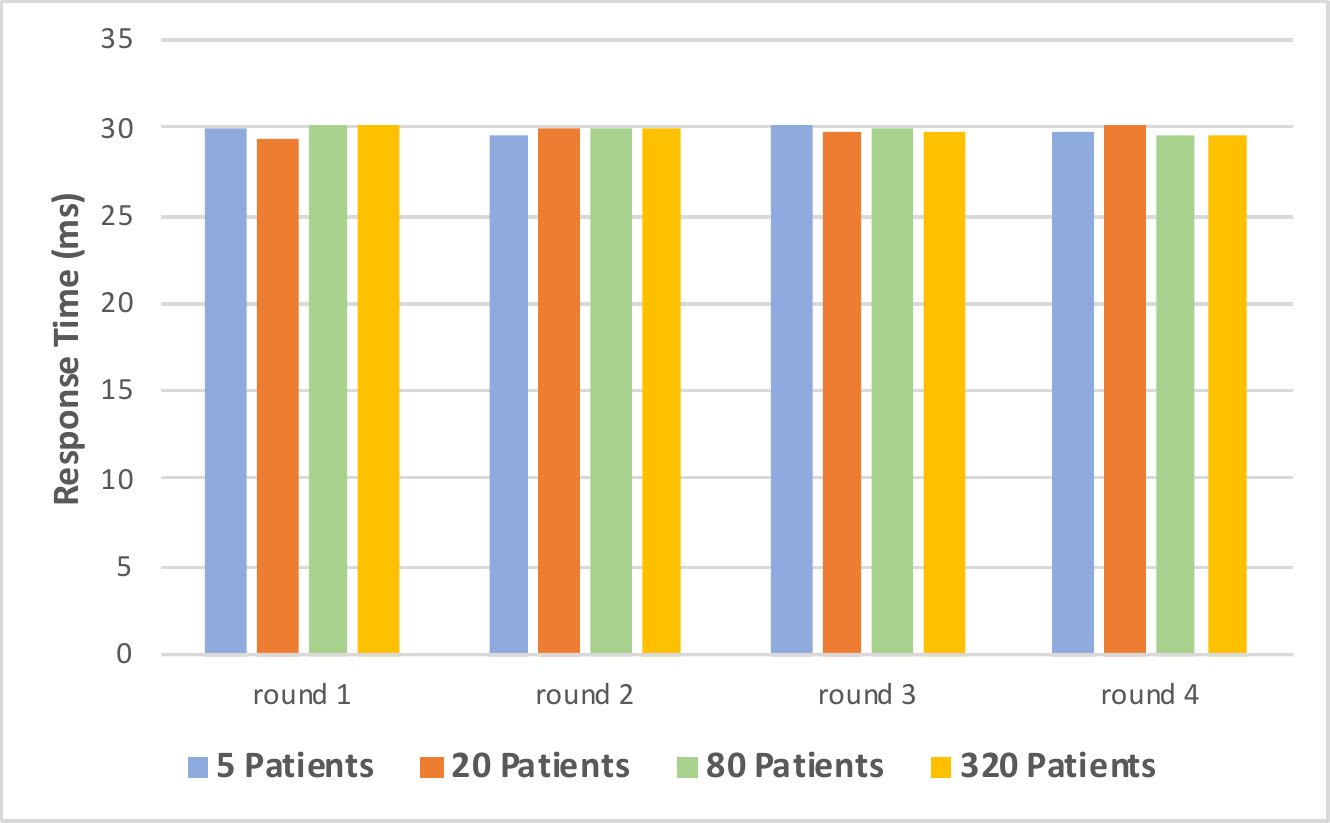}
\caption{Response time against unauthorized retrievals.}
\label{fig:responsetime}
\end{figure}

\begin{figure}[h]
\centering
\includegraphics[width=0.415\textwidth]{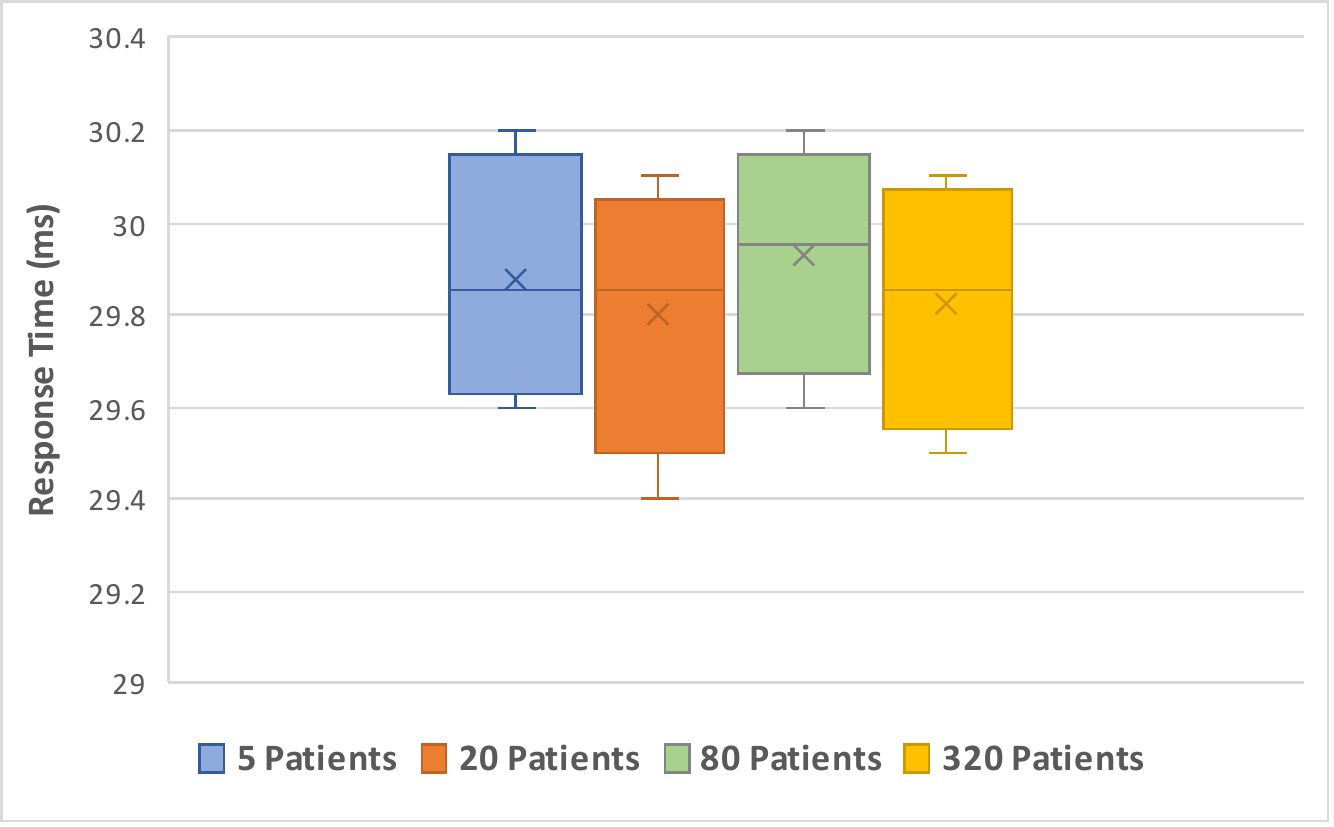}
\caption{Average response time against unauthorized retrieval.}
\label{fig:responsetime_box}
\end{figure}

To evaluate the scalability of our design, we also conduct extensive system performance evaluation by increasing the number of participants from initially 5 patients to 20 patients, 80 patients and 320 patients. The average transaction processing time from 4 rounds of tests [Fig. \ref{fig:transactiontime}] remains at around 40 ms as shown in Fig. \ref{fig:transactiontime_box}. 
The average response time against unauthorized retrieval from 4 rounds of  tests [Fig. \ref{fig:responsetime}] also remains at around 30 ms as shown in Fig. \ref{fig:responsetime_box}. The results of the extensive experiments demonstrate that the performance of the proposed system is scalable.

\section{Security Analysis }
\label{security}

In this section, we analyze the robustness of the proposed blockchain-based access control scheme against the potential attacks.

\subsection{EHR spoofing attack}
In the proposed system, EHR data is remotely stored on an edge node. In the meanwhile, the hash digest result about EHR data has been stored permanently on the blockchain network with all records of access control log. When a malicious user attempts to tamper the EHR data,  the newly generated hash digest will be different from the original hash digest stored in the blockchain. As a result, participants immediately find that the EHR data has been attacked. Moreover, the proposed scheme enforces ABAC policies on EHR data which will also protect the integrity and security of EHR data since unauthorized user cannot obtain the access permission.

\subsection{Fake healthcare provider attack}
As mentioned previously, our proposed scheme utilizes the Hyperledger Composer Fabric blockchain platform as the external controller of the blockchain network, which manages the identities and access control log by various events and transactions. The healthcare providers are required to get a unique ID by registering in the blockchain. Without the provider's complete registry information, a malicious user cannot impersonate the provider in order to request the EHR data. If a user tries to retrieve information via malicious attacks, the system can blacklist the account and all the actions will be recorded as evidences. If all the personal identity information along with attributes-related information belonging to a provider has been leaked to a malicious user, he or she may access the EHR data. However, all the access requests and actions will have been permanently recorded in the blockchain. When it is found later that this user's account has been hacked, the system can trace back the attacker's actions and identify the changed data.



\section{Related Work}
\label{related}

In this section, we review the related work on blockchain-based access control. Azaria et al.~\cite{azaria2016medrec} proposed a decentralized record management system to handle EHRs. The system gives patients an immutable log and easy access to their medical information. However, the protocol used is based on consensus mechanism of proof-of-work which consumes massive computing resources.  Measa et al.~\cite{maesa2017blockchain} proposed a blockchain-based solution to publishing the policies expressing access rights of resources and allowing the distributed transfer of such rights among users. The authors presented a Bitcoin-based proof-of-concept implementation without describing any experiment or evaluation. Wang et al.~\cite{yue2016healthcare} proposed an iOS App termed Healthcare Data Gateway based on blockchain to enable patient to own, control and share their own data. Ekblaw et al.~\cite{ekblaw2016case} proposed the MedRec prototype to give patients an immutable log and access to their medical information across healthcare providers and treatment sites. They proposed that medical stakeholders can participate in the network as blockchain ``miner.'' Guo et al.~\cite{guo2018secure} proposed an attribute-based signature scheme with multiple authorities for EHR data. In this scheme, a patient endorses a message according to her attributes while disclosing no information other than the evidence that the patient has attested to it. Xia et al.~\cite{xia2017medshare} proposed the MedShare, a blockchain-based system for sharing medical data in cloud repositories among different entities. They deploy smart contracts and access control mechanisms to track the dataflow and to detect violations.

There have been various attempts to address the proper access control issues on data management using blockchain. Zyskind et al.~\cite{zyskind2015decentralizing} described a decentralized data
management system which ensures users own and control their
data and proposed a protocol to enable automated access-control manager using multi-party computation. Wang et al.~\cite{wang2018blockchain} described a blockchain-based framework for data sharing in decentralized storage systems combining Ethereum blockchain and attribute-based encryption technology. Guo et al.~\cite{guo2019multi} proposed an multi-authority attribute-based access control mechanism by utilizing Ethereum’s smart contracts. Ouaddah et al.~\cite{ouaddah2016fairaccess} proposed the FairAccess framework to include transactions used to grant, get, delegate, and revoke access. As a proof of concept, FairAccess was implemented on a Raspberry Pi device using a local blockchain.

To the best of our knowledge, our work is the first study to integrate both the Hyperledger Composer Fabric blockchain platform providing the on-chain access control logs for EHR data with smart contracts and ACL, and the off-chain edge nodes to enforce ABAC policies on EHR data. We also developed a prototype of blockchain network to evaluate the system performance.

\section{Conclusion}
\label{conclusion}


In this paper, we propose a hybrid architecture of using both blockchain and edge nodes to impose attribute-based access control of EHR data. The architecture utilizes blockchain (1) to execute smart contracts so as to impose ACL policy and (2) to record legitimate access events into blockchain. In addition, EHR data is stored on edge nodes which impose ABAC policies specified in ALFA. We used the Hyperledger Composer Fabric blockchain programmed with smart contracts and ACL policies to evaluate the performance of access control by measuring the transactions processing time and response time against unauthorized retrieval attempts. The experiments show that our system provides results in terms of milliseconds making it suitable to be incorporated in real-time and secured EHR data access control frameworks. For future work, we plan to investigate novel consensus protocol designs for the proposed mechanism to achieve better performance. In addition, we plan to develop a Hyperledger-based benchmark tool for a set of performance analyses.





%

\bibliographystyle{IEEEtran}
\bibliography{bibtex/IEEEabrv,bibtex/sigproc}

\end{document}